\documentclass[aip,jcp,reprint]{revtex4-1}
\pdfoutput=1

\newcommand{\beginsupplement}{%
    \setcounter{equation}{0}
    \renewcommand{\theequation}{S\arabic{equation}}
    \setcounter{section}{0}
    \renewcommand{\thesection}{S\arabic{section}}
    \setcounter{table}{0}
    \renewcommand{\thetable}{S\arabic{table}}%
    \setcounter{figure}{0}
    \renewcommand{\thefigure}{S\arabic{figure}}%
    }

\usepackage{amsmath,amssymb}
\usepackage{graphicx}
\usepackage{hyperref}

\newcommand{\unif}{\operatorname{unif}}
\newcommand{\ang}[1]{\operatorname{ang_#1}}
\newcommand{\dirf}{\operatorname{dir}}
\newcommand{\cb}{\tilde{\chi}}
\newcommand{\Vb}{\tilde{V}}
\newcommand{\Vt}{\bar{V}}
\newcommand{\single}{^{(1)}}
\newcommand{\pair}{^{(2)}}

\begin{document}

\title{ Rapid calculation of side chain packing and free energy with
applications to protein molecular dynamics}

\author{J.M.~Jumper}
\email[]{jumper@uchicago.edu}
\affiliation{Department of Chemistry, University of Chicago}

\author{K.F.~Freed}
\email[]{freed@uchicago.edu}
\affiliation{Department of Chemistry, James Franck Institute, University of Chicago}

\author{T.R.~Sosnick}
\email[]{trsosnic@uchicago.edu}
\affiliation{Department of Biochemistry and Molecular Biology, Institute for Biophysical Dynamics, University of Chicago}

\date{\today}

\begin{abstract}
    To address the large gap between time scales that can be easily
    reached by molecular simulations and those required to understand protein
    dynamics, we propose a rapid  self-consistent
    approximation of the side chain free energy at every integration step.  In
    analogy with the adiabatic Born-Oppenheimer approximation for electronic structure,
    the protein
    backbone dynamics are simulated as preceding according to the dictates of
    the free energy of an instantaneously-equilibrated side chain potential.
    The side chain free energy is computed on the fly, allowing the protein
    backbone dynamics to traverse a greatly smoothed energetic landscape.  This results
    in extremely rapid equilibration and sampling of the Boltzmann distribution.
    Because our method employs a reduced model involving single-bead side
    chains, we also provide a novel, maximum-likelihood method to
    parameterize the side chain model using input data from high resolution
    protein crystal structures.  We demonstrate state-of-the-art accuracy for
    predicting $\chi_1$ rotamer states while consuming only milliseconds of CPU
    time.  We also show that the resulting free energies of side chains is
    sufficiently accurate for \textit{de novo} folding of some proteins.
\end{abstract}

\maketitle

\section{Introduction} Two major challenges must be overcome in order to
accurately simulate protein dynamics.  The first is the necessity of balancing
the large and competing sources of energy and entropy whose sum determines both
the thermodynamics and the native conformation of the protein.  The second
challenge involves the intensive sampling required to obtain a Boltzmann
ensemble of conformations.  The sampling challenge is addressed here by
integrating out the side chain degrees of freedom to produce a coarse-grained
configuration defined just in terms of the backbone N, C$_\alpha$, and C atoms.
Consequently, backbone motions evolve on a smoother coarse-grained free energy
surface with greatly reduced side chain rattling (molecular friction) compared
to that for standard all-atom molecular dynamics simulations.

The uncertainty in the position of coarse-grain interactions heightens the
difficulty of accurately parameterizing a coarse-grained model to represent the
physical interactions. Moreover, all-atom force fields produce conformations
that deviate from experiment, especially for unfolded
proteins\cite{skinner2014benchmarking}.  We do not follow the customary process
of matching the energies of the coarse-grained model to approximate the already
inexact energies of atomistic force fields or try to interpret raw statistics
for the distribution of interatomic distances in the Protein Data Bank
(PDB)\cite{berman2000protein} through a reference state
\cite{hamelryck2010potentials}.  Instead, our side chain energies are determined
as those that best reproduce the side chain conformations observed in the PDB
when using the native-state backbone configurations.

This maximum-likelihood approach has key advantages: (1) it directly provides an
interpretation of the structural information as a sample from the statistical
mechanical ensemble of side chain packing, and (2) it can be evaluated quickly
since we show that approximating the Boltzmann distribution for the side chains
in a fixed backbone configuration does not require laborious discrete sampling
of the $\chi$ angles in the side chain.  Our method enables rapidly
equilibrating coarse-grained simulation that can nonetheless contain significant
molecular detail.

While the overarching goal of our work is preparation for extremely rapid
molecular dynamics, our new interaction model gives very accurate and rapid
predictions of side chain $\chi_1$ angles.  Using our side chain ensembles, we
are able to predict $\chi_1$ rotamer configurations with accuracy exceeding the
state of the art from SCWRL4\cite{scwrl4}, yet our predictions take less than
1\% of the computational time.  We are also exceed the performance of the rapid
side chain packing algorithm RASP\cite{rasp} by more than an order of magnitude.
The accuracy of our side chain rotamer predictions validates that our side chain
interaction potential captures important physics for side chain interactions,
suggesting suitability for molecular dynamics.

The software for our side chain packing and molecular dynamics methods is open
source.  The source code may be obtained at
\url{https://github.com/John-Jumper/Upside-MD}, where the version tagged
\texttt{sidechain\_paper} should be used to reproduce the results of this paper.

\section{Upside Model and  side chain free energy evaluation}

\begin{figure*}
    \centering
    \includegraphics[width=0.8\textwidth]{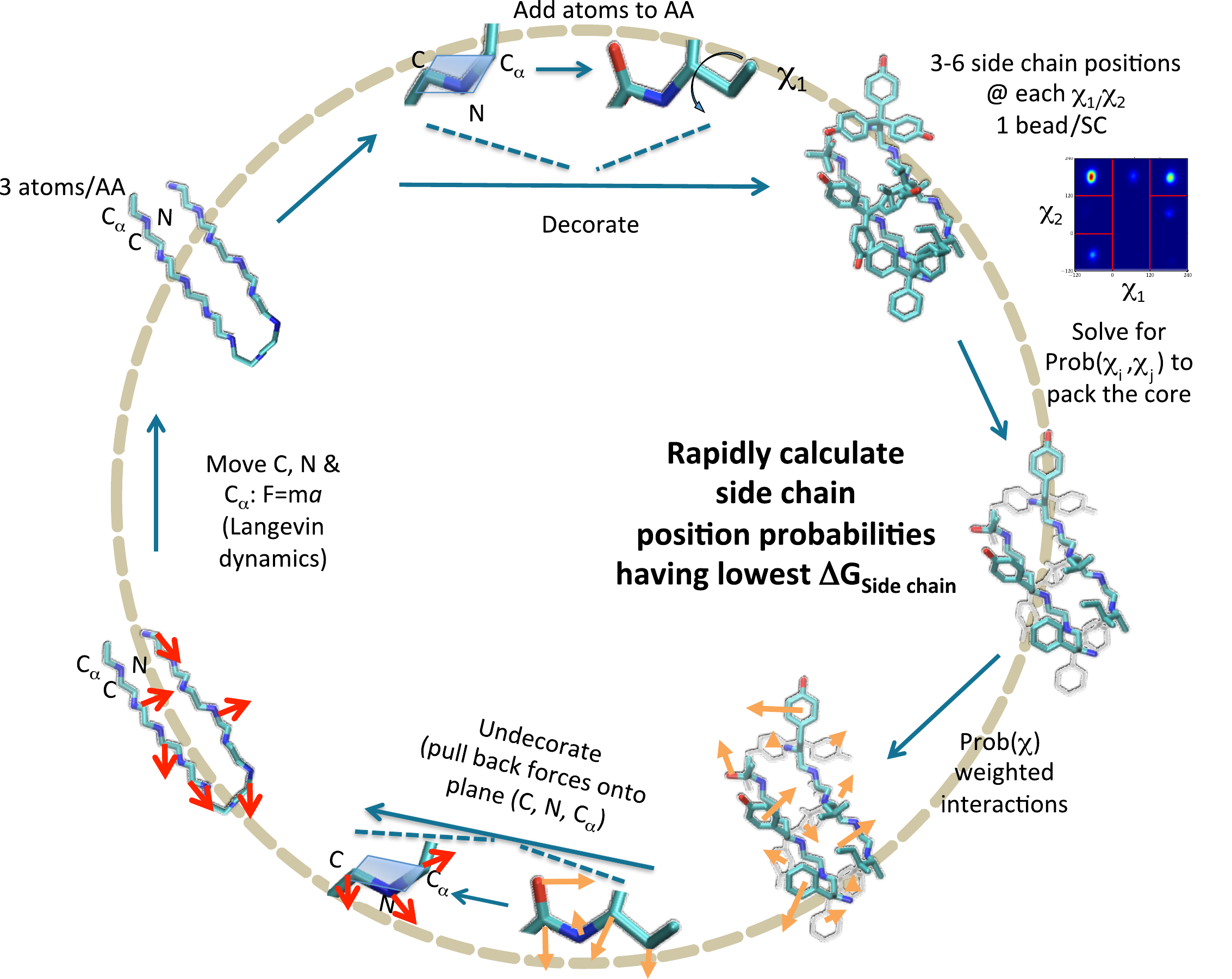}
    \caption{
	\label{fig:comp_loop} Inner loop of Upside calculation.  The  
        side chain potential enters into the integration step simply
        as a complicated, many-body energy function that may be treated with 
        standard techniques of molecular simulations.} 
\end{figure*}

The positions of the  N, C$_\alpha$, and C atoms constitute the backbone trace.
The trace determines the fold of the protein, and the free energy of the trace
is likely to preserve the major barriers that determine the long-timescale
degrees of freedom in the protein. The strategy in our Upside model is to
perform dynamics simulations of the backbone trace, while still including
sufficient structural details (side chain structures and free energies, etc.)
necessary to compute realistic forces on the three atoms of the backbone trace.
The inclusion of the side chain free energy, rather than the side chains
themselves, greatly smooths the potential governing the dynamics of the backbone
trace, especially because of the reduction of steric rattling.

One can consider a representation of the protein configurations in terms of the
coordinates $(\{b_i\}, \{\chi_i\})$ where $b_i$ represents the positions of the
backbone N, C$_\alpha$, and C atoms on the $i$-th residue and $\chi_i$
represents the side chain $\chi$-angles on the $i$-th residue.  Since bond
lengths and angles are approximately constant for proteins, the positions of the
protein atoms can be reconstructed with high accuracy from the $(\{b_i\},
\{\chi_i\})$ coordinates.  Given a potential energy $V(\{b_i\}, \{\chi_i\})$, we
define the free energy as a function of the backbone configuration,
\begin{align}
    \Vt(\{b_i\}) = -\log \int d\chi_1\,\cdots\chi_N\, e^{-V(\{b_i\}, \{\chi_i\})}.
    \label{eqn:vt-defn1}
\end{align}
Natural energy units are used so that $k_\text{B}T = 1$.  An intermediate step
of this derivation requires the introduction of a discrete approximation
$\{\cb_i\}$ for our $\chi$-angles and a discrete approximation $\bar{V}(\{b_i\},
\{\cb_i\})$ for the potential. 

\begin{figure}
\centering
    \includegraphics[width=3.5in,trim={0.5in 0 0 0 },clip]{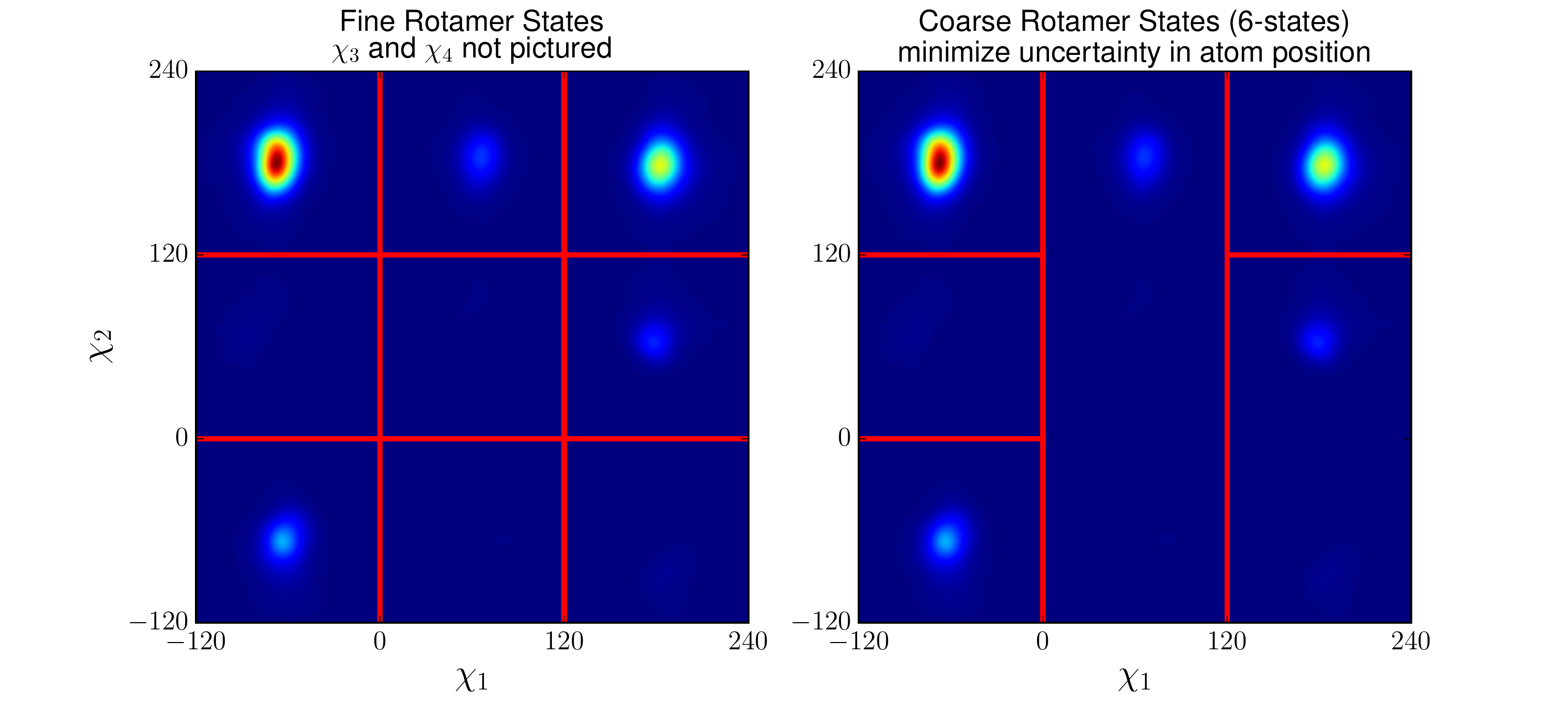}
    \caption{\label{fig:coarse-states} Example of optimized coarse states for
    arginine overlaid on the PDB distribution of the rotamer angles $\chi_1$ and
    $\chi_2$.  Each of the six coarse states contains only a single fine state
    that has high probability, so that the variance of dihedral angles within
    each coarse state is small.}
\end{figure}

Rather than directly calculate Eq.~\ref{eqn:vt-defn1}, we define an
intermediate discrete approximation to $\Vt$ that is amenable to approximation
techniques.  Consider a discrete coarse-graining function $g$ so that $\cb_i =
g(\chi_i)$, where $\cb_i$ is a state label ($\cb_i \in \{1,\ldots,6\}$ as each
side chain is represented by a bead located at one of up to 6 positions). The
coarse-grain potential $\Vb$ is defined so that 
\begin{align}
    e^{-\Vb(\{b_i\}, \{\cb_i\})} \approx  \int d\chi_1\,\cdots\chi_N\, \left(\prod_i
    \delta_{\cb_i f(\chi_i)}\right) e^{-V(\{b_i\}, \{\chi_i\})}.
    \label{eqn:vb-defn}
\end{align}
In principle, any coarse-grain function for the side chains may be used.  The
discrete form $\Vb$ to the potential provides an accurate approximation as the
distribution of $\chi$-angles is sharply peaked (in the true potential $V$)
within each discrete state $\cb$.  See Figure \ref{fig:coarse-states} for an
example of a function and section \ref{sec:state-decomposition} where an
optimized function $f$ is derived.  We make the following assumptions on the
form of $\Vb$.  First, we assume an explicit function $y_i(b_i, \cb_i)$ exists
for the side chain coordinates based only on the backbone coordinates and side
chain state for residue $i$.  We may relax the requirement to consider a single
residue's backbone position, but it is required that $y_i$ depend on only a
single side chain \textit{state} $\cb_i$.  These directed coordinates are
approximately side chain centers of mass with direction given by the
C$_\beta$--C$_\gamma$ bond vector.  A further assumption is that $\Vb$ can be
expressed in the form
\begin{align}
    \Vb(\{b_i\}, \{\cb_i\}) = V^\text{backbone}(\{b_k\}) + \nonumber \\
    \sum_i V\single_i(\{b_k\}, \cb_i, y_i(b_i,\cb_i)) +  \nonumber\\
    \sum_{i,j} V\pair_{ij}(y_i(b_i,\cb_i), y_j(b_j,\cb_j)),
\end{align}
where the  pair interaction $V\pair_{ij}(y_i,y_j) = 0$ for the side chain is
taken to vanish beyond a cutoff $R_\text{cutoff}$.  The dependence of the
potential on the backbone is completely general, but the potential is assumed to
contain at most a pairwise dependence on the discrete rotamer states $\cb_i$.
Explicit parameterizations for $y_i$ and $\Vb$ are defined in Section
\ref{sec:parametric} using the principle of maximum likelihood.

One can simulate the Boltzmann ensemble for $\Vb$ using molecular dynamics for
the backbone $\{b_i\}$ and Monte Carlo moves for the side chain states
$\{\cb_i\}$.  But the strong steric interactions are likely to lead to slow
equilibration and dynamics for both the side chains and backbone. Because we are
predominantly interested in backbone motions, we return to the free energy $\Vt$
in Eq.~\ref{eqn:vt-defn1}, now summing over discrete side chain states instead
of integrating over continuous side chain angles,
\begin{align}
    e^{-\Vt(\{b_i\})} \approx \sum_{\cb_1,\ldots,\cb_N} e^{-\Vb(\{b_i\}, \{\cb_i\})}.
    \label{eqn:vt-defn2}
\end{align}
The potential $\Vt$ represents a further coarse-graining of the system by
completely replacing the influence of the side chains with a potential
describing their adiabatic free energy for a given fixed backbone conformation.
Because $\Vt$ depends only on the (continuous) backbone coordinates, this choice
of $\Vt$ enables running standard molecular dynamics simulations instead of a
hybrid of Monte Carlo and molecular dynamics.  

Importantly, the potential $\Vt$ is a much smoother function of the backbone
coordinates than the original $V(\{b_i\},\{\chi_i\})$ because the replacement of
the side chain degrees of freedom with the approximate free energy of the side
chains greatly reduces steric rattling and molecular friction. The reduction of
the ruggedness of the energy landscape enhances diffusion within conformational
basins but preserves the overall structure and barriers that define the
conformational ensemble.

\section{Approximating the side chain free energies}

The benefits of running dynamics with the coarse grained $\Vt$ enter at great
cost because using even three coarse-grained states per side chain implies a
summation over $3^N$ $\cb$-states in Eq.~\ref{eqn:vt-defn2}.  Furthermore, the
vast majority of those $3^N$ states have steric clashes or other large energies
and, therefore, contribute little to the free energy of the side groups.  

To approximate the free energy of the side chains $\Vt$, we express the problem
in the language of Ising models so that we can apply standard techniques
developed in that context.  For a fixed backbone configuration $\{b_i\}$, 
\begin{align}
    \Vb(\{b_i\}, \{\cb_i\}) &= \bar{v}(\{\cb_i\}) \nonumber \\
    &= \sum_i v\single_i(\cb_i) + \sum_{\substack{i,j\\ \text{neighbors}}}v\pair_{ij}(\cb_i, \cb_j),
    \label{eqn:ising}
\end{align}
where the potentials $\bar{v}$ are written in lowercase to indicate  suppression
of the dependence on the fixed backbone coordinates $\{b_i\}$ in order to focus
on the side chain contribution.  Notice that with the backbone positions fixed,
each single-residue potential $v\single_i$ is simply a vector with as many
components as the number of possible states for $\cb_i$ (e.g.~length-6 vectors).
Similarly, each of the pair potentials $v\pair_{ij}$ is a small 6x6 matrix of
potential energies to cover a maximum of 36 possibilities. These single and pair
potentials are calculated only once before evaluating the free energy as
described in section \ref{sec:parametric}.  Moreover, the pair summation in
Eq.~\ref{eqn:ising} only applies for residues pairs $i$ and $j$ that are
neighbors spatially.  A pair of residues $(i,j)$ are neighbors if inter-residue
distance $|y_i(\cb_i) - y_j(\cb_j)|$ is less than a cutoff $R_\text{cutoff}$ for
any of their possible discrete states $(\cb_i,\cb_j)$. In this work, we use
$R_\text{cutoff} = 7\text{\AA}$ for side chain-side chain interactions and
$R_\text{cutoff} = 5\text{\AA}$ for side chain-backbone interactions.

\begin{figure}
\centering
    \includegraphics[width=1.5in]{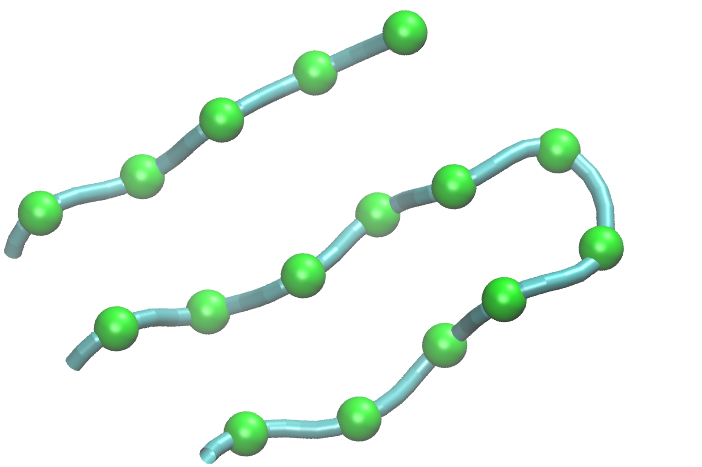}
    \includegraphics[width=1.5in]{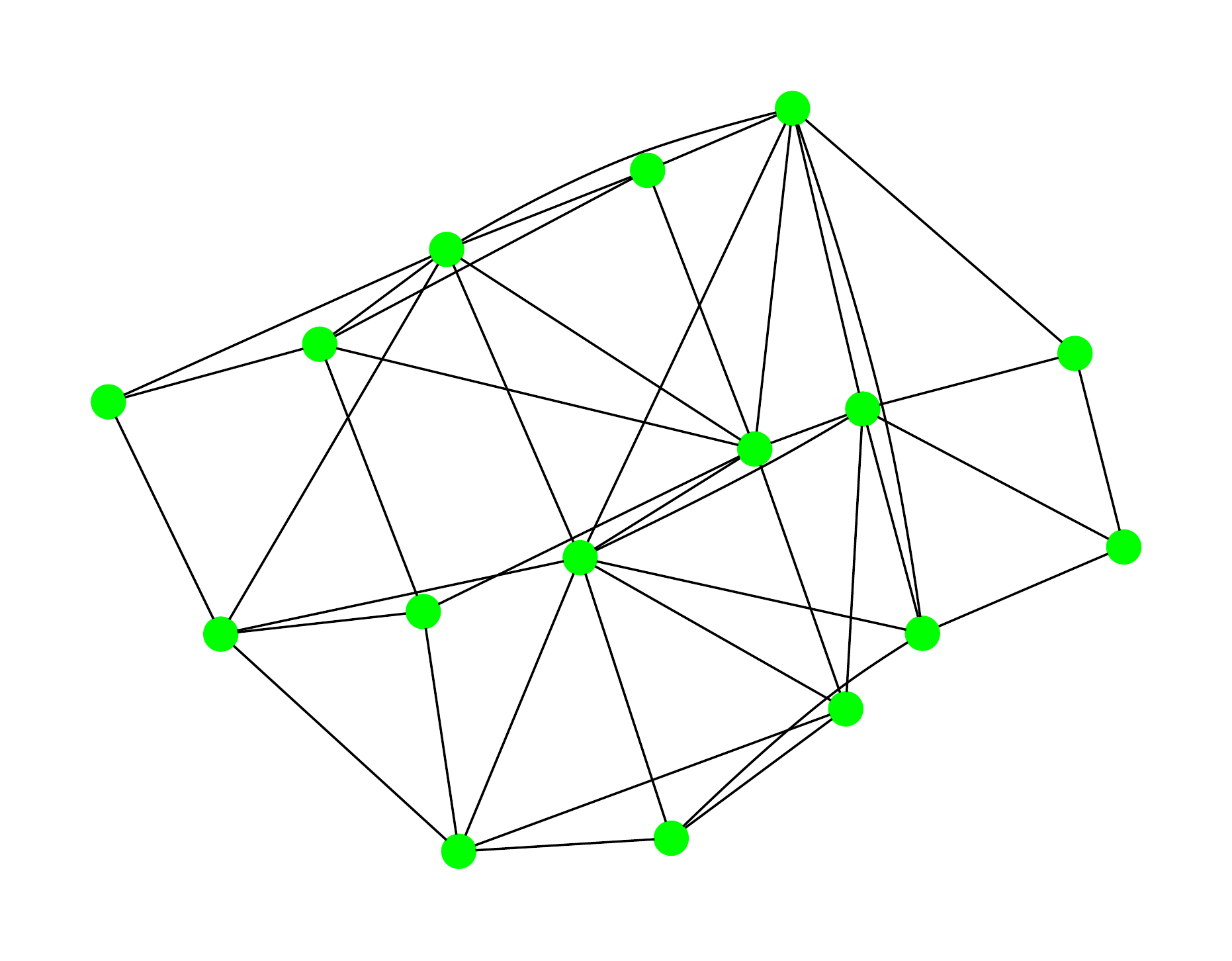}
    \caption{\label{fig:ising}
    Fragment of protein G with associated interaction graph ($R_\text{cutoff}
    = 7\text{\AA}$).  A pair of residues has a connection whenever their side
    chain beads are within $R_\text{cutoff}$ for any side chain states.}
\end{figure}

The potential $\Vb$ may be visualized as an energy function on a graph with one
discrete site per amino acid.  The graph has a connection between any two
residues that are within the cutoff separation $R_\text{cutoff}$ (Figure
\ref{fig:ising}). The structure of this graph varies dynamically over the course
of a simulation because the definition of neighboring residues depends on the
backbone configuration $\{b_i\}$.  The potential varies smoothly as the backbone
moves so long as the pairwise potential functions are continuous in the backbone
coordinates.  The potential $\Vb$ is continuous despite the changing connections
of the graph because the strength of the potential for each interaction
approaches zero at $R_\text{cutoff}$ just before the connection is eliminated
from the graph.  Problems such as this, with discrete potentials on an arbitrary
graph, are extensively studied in both statistical mechanics (as variants of the
Ising model) and machine learning (as undirected graphical models or Markov
random fields)\cite{wainwright2008graphical}.  Below we adopt some well studied
approximations from these fields to provide accurate and tractable methods for
computing our coarse-grain potential $\Vt$.

Two approximations (see reference \citenum{wainwright2008graphical}) are invoked
to compute the free energy from
\begin{align}
    \Vt = G^\text{SC} = -\log \sum_{\cb_1,\ldots,\cb_N} e^{-v(\{\cb_i\})}.
\end{align}
The first approximation is to express the free energy $G^\text{SC}$ in terms of
the entropy and average energy of the Boltzmann ensemble where the entropy has
been replaced by an approximation,
\begin{align}
    G^\text{SC} &=       \langle \bar{v} \rangle - S \nonumber \\
		&\approx \langle \bar{v} \rangle - S^\text{approx}, \label{eqn:G-approx}
\end{align}
where $\langle \bar{v} \rangle$ and $S^\text{approx}$ are defined in 
Eq.~\ref{eqn:barv} and Eq.~\ref{eqn:sapprox}.  We express the average energy and
approximate entropy using the single-residue probabilities $p_i(\cb_i)$ that
residue $i$ is in state $\cb_i$ in the Boltzmann ensemble of $\bar{v}$ and
similarly for the joint probabilities $p_{ij}(\cb_i,\cb_j)$.  Using $p_i$ and
$p_{ij}$, the approximate energy and entropy are
\begin{align}
    \langle \bar{v} \rangle = &\sum_i \sum_{\cb_i} p_i(\cb_i) v\single_i(\cb_i) + \nonumber \\
    &\sum_{\substack{i,j\\ \text{neighbors}}}\sum_{\cb_i,\cb_j} p_{ij}(\cb_i,\cb_j)
    v\pair_{ij}(\cb_i, \cb_j) \label{eqn:barv}\\
    \nonumber \\
    S^\text{approx} = &\sum_i \sum_{\cb_i} p_i(\cb_i) (-\log p_i(\cb_i)) - \nonumber \\
    &\sum_{\substack{i,j\\ \text{neighbors}}}\sum_{\cb_i,\cb_j} p_{ij}(\cb_i,\cb_j)\log
    \frac{p_{ij}(\cb_i,\cb_j)}{p_i(\cb_i),p_j(\cb_j)}. \label{eqn:sapprox}
\end{align}           
The mutual information approximation to the entropy ignores contributions from
three-residue and higher correlations.  

We intend to minimize the approximate free energy in
Eq.~\ref{eqn:G-approx} over all putative Boltzmann probability distributions for
the side chain states $\{\cb_i\}$.  Notice that only the 1-side chain
probabilities $p_i$ and 2-side chain probabilities $p_{ij}$ are required to
compute the average energy and approximate entropy; we do not need the more
complicated full joint probability distribution of the $\{\cb_i\}$ states for
all side chains.  In addition to the mutual information approximation of the
entropy, we assume that any pair probability $p_{ij}$ represents possible pair
probabilities from a Boltzmann distribution, so that the only task is to
minimize the free energy with respect to the pair probabilities.  The only
constraints imposed are that they must satisfy the obvious consistency
conditions for probabilities,
\begin{eqnarray}
    p_j(\cb_j) = \sum_{\cb_i} p_{ij}(\cb_i) = \sum_{\cb_k} p_{jk}(\cb_k)  \label{eqn:first-cond} \\
    \sum_{\cb_i,\cb_j} p_{ij}(\cb_i,\cb_j) = 1 \\
    p_{ij}(\cb_i,\cb_j) = p_{ji}(\cb_j,\cb_i). \label{eqn:last-cond}
\end{eqnarray}
However, use of only conditions
in Eq.~\ref{eqn:first-cond}--\ref{eqn:last-cond} is insufficient to ensure that a
joint probability distribution exists for all the variables consistent with
the choices of $p_i$ and $p_{ij}$.  As an explicit example, 
\begin{align}
    p_{12} &= p_{23} = \begin{pmatrix} 1/3 & 0 & 0 \\ 0 & 1/3 & 0 \\ 0 & 0 & 1/3 \end{pmatrix} \\
    p_{13} &= \begin{pmatrix} 1/9 & 1/9 & 1/9 \\ 1/9 & 1/9 & 1/9 \\ 1/9 & 1/9 & 1/9 \end{pmatrix}
\end{align}
obeys conditions Eq.~\ref{eqn:first-cond}--\ref{eqn:last-cond} but is not
representable by any probability distribution on the three residues.  This
aspect is a result of residue 1 being completely correlated to residue 2, and
residue 2 being completely correlated to residue 3, but residues 1 and 3 being
independent, which is mathematically impossible.  Though this
non-representability problem is a potential concern, we expect that it will not
be a large source of error.

Accepting the two approximations for entropy and representability, the free
energy becomes
 \begin{align}
     G^\text{SC} \approx \min_{\{p_i\},\{p_{ij}\}} (\langle \bar{v} \rangle - S^\text{approx}). \label{eqn:free-energy}
 \end{align}
Thus, we now have a tractable approximation to free energy of the side chain. We
can minimize that free energy using a self-consistent iteration technique called
belief propagation (\ref{apx:beliefs}).  The iteration typically
converges rapidly, often in 10-20 steps.

Molecular dynamics simulations require calculations of the forces on the
backbone coordinates, $-\frac{d\Vt}{db_i}$.  The forces are obtained from the
derivatives of the potential computed using the chain rule.  We take advantage
of several terms being zero because the pair probabilities minimize the free
energy,
\begin{align}
    \frac{d G^\text{SC}}{db_k} &= \frac{\partial G^\text{SC}}{\partial b_k} +
    \sum_i \frac{\partial G^\text{SC}}{\partial p_i} \frac{\partial p_i}{\partial b_k} +
    \sum_{\substack{i,j\\ \text{neighbors}}} \frac{\partial G^\text{SC}}{\partial p_{ij}}
    \frac{\partial p_{ij}}{\partial b_k}  \nonumber \nonumber \\
    &= \frac{\partial G^\text{SC}}{\partial b_k} 
    = \frac{\partial \langle \bar{v} \rangle}{\partial b_k}
    = \left\langle \frac{\partial \bar{v}}{\partial b_k} \right \rangle \nonumber \\
    &=\sum_i \sum_{\cb_i} p_i(\cb_i) \frac{\partial v\single_i}{\partial b_k}(\cb_i) + \nonumber \\
    &\hphantom{=}\sum_{\substack{i,j\\ \text{neighbors}}}\sum_{\cb_i,\cb_j} p_{ij}(\cb_i,\cb_j)
    \frac{\partial v\pair_{ij}}{\partial b_k}(\cb_i, \cb_j) \label{eqn:ref-param-deriv}
\end{align}
where $\frac{\partial G^\text{SC}}{\partial p_i} = \frac{\partial
G^\text{SC}}{\partial p_{ij}}=0$ because $p_i$ and $p_{ij}$ are chosen to
minimize $G^\text{SC}$.  The remaining simplifications occur because
$S^\text{approx}$ is independent of the backbone coordinates.  While the
underlying side chain interactions are pairwise additive and vanish outside the
cutoff radius $R_\text{cutoff}$, the free energy in equation
Eq.~\ref{eqn:G-approx} is a many-body potential that can interact over arbitrary
distances.

Since the approximate free energy due to the side chains is not a convex
function of the probabilities, local minima may arise and impair the
self-consistent iteration from finding the global minimum.  To reduce the danger
posed by the presence of local minima, calculations are begun from a carefully
initialized state (see \ref{apx:beliefs} for details). Other
self-consistent approximations exist for the side group free energy, such as
tree-reweighted belief propagation\cite{wainwright2003tree}, that are typically
less accurate but always converge to the global minimum of their approximate
free energy.  Another limitation of the present approximation scheme arises when
a bi-stable or multi-stable energy landscape is possible for the rotamer states.
If well-separated and equally important minima are present for a single backbone
configuration in the rotamer free energy surface, the probabilities only
converge to a single minimum and thus underestimate the entropy of the side
chains. While this does not appear to occur near the native well, we have not
extensively searched for special backbone configurations that would result in
bi-stable rotamer energies.  The characterization of such problematic
configurations, likely near free energy barriers, is left to future work.  

\section{Bead locations and interactions} \label{sec:parametric} 
	
\begin{figure*}
\begin{minipage}[c]{4in} 
\vspace{0pt}
\centering
    \includegraphics[width=3.5in]{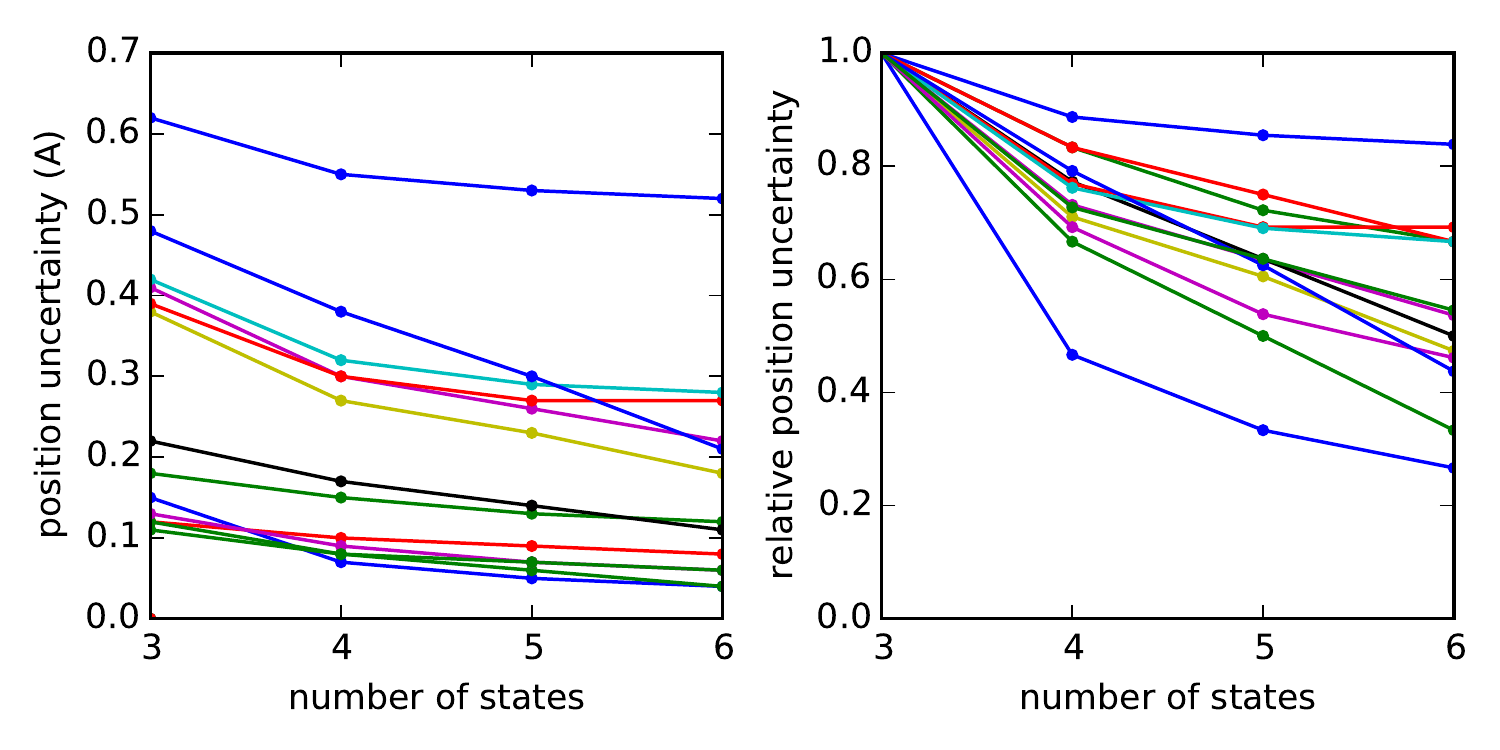}
\end{minipage}
\begin{minipage}[c]{2in} 
\vspace{0pt}
\centering
    \begin{tabular}{llll}
    Restype &  States & Restype & States  \\
        ALA &         1 & LEU &         3 \\
        ARG &         6 & LYS &         3 \\
        ASN &         6 & MET &         6 \\
        ASP &         6 & PHE &         6 \\
        CYS &         3 & PRO &         3 \\
        GLN &         6 & SER &         3 \\
        GLU &         6 & THR &         3 \\
        GLY &         1 & TRP &         6 \\
        HIS &         6 & TYR &         6 \\
        ILE &         3 & VAL &         3 \\
        \end{tabular}
\end{minipage}
    \caption{Error in the decomposition of rotamer states into coarse-grained
    states as a function of the number of side chain states.  The table
    summarizes the number of states chosen for each amino acid type. The
    uncertainty in position is $\sigma$.  The relative uncertainty is the
    position uncertainty for each number of states divided by the accuracy at
    three states. One, three or six rotamer states are used, depending on the
    residue type. For residues without a rotatable $\chi_2$, such as valine,
    only three states are needed. The computational time to compute the pairwise
    interactions and  solve for the free energy scales roughly as the number of
    coarse rotamer states squared, so the use as fewer coarse states is
    preferred.}
\end{figure*}

Paralleling the necessity of coarse-graining the rotamer states, side chain
atoms themselves also require coarse-graining in order to obtain an inexpensive
side chain model.  This reduction in the number of degrees of freedom is further
justified since the atomic positions of the side chains are uncertain due to the
discretization and aggregation of the rotamer states, meaning that there is
little value in assigning precise positions for all atoms.  We instead use a
single oriented bead (location and direction coordinates) to represent each side
chain (note that the direction is independent of the side chain, e.g.~in
aromatic residues it could be the ring normal unit vector). The locations and
directions of the side chain beads are changed by the optimizer during the
optimization of the potential.  The improvement in prediction accuracy from
using optimized side chain positions rather than the static positions is
surprisingly substantial.

We use a combination of isotropic and directional interactions for each pair of
interacting side chain or backbone types.  The isotropic interactions are
primarily responsible for enforcing excluded volume, while the directional
interactions typically reflect specific chemical interactions arising from polar
groups or aromatic ring stacking.  Concretely, each interaction pair is
described by positions $y_1$ and $y_2$ and directions $n_1$ and $n_2$.  From
this the distance $r_{12} = |y_1 - y_2|$ and displacement unit vector $n_{12} =
(y_1-y_2)/r_{12}$ are calculated.  The form of the interaction is given by 
\begin{align}
    V = \kappa (&\unif(r_{12}) + \nonumber \\
		       &\ang{1}(-n_1 \cdot n_{12}) \ang{2}(n_2\cdot n_{12}) \dirf(r_{12})),
		       \label{eqn:interaction}
\end{align}
where $\unif$, $\ang{1}$, $\ang{2}$, and $\dirf$ are smooth curves represented
by cubic splines.  

\begin{figure}
    \centering
    \includegraphics[width=2.5in]{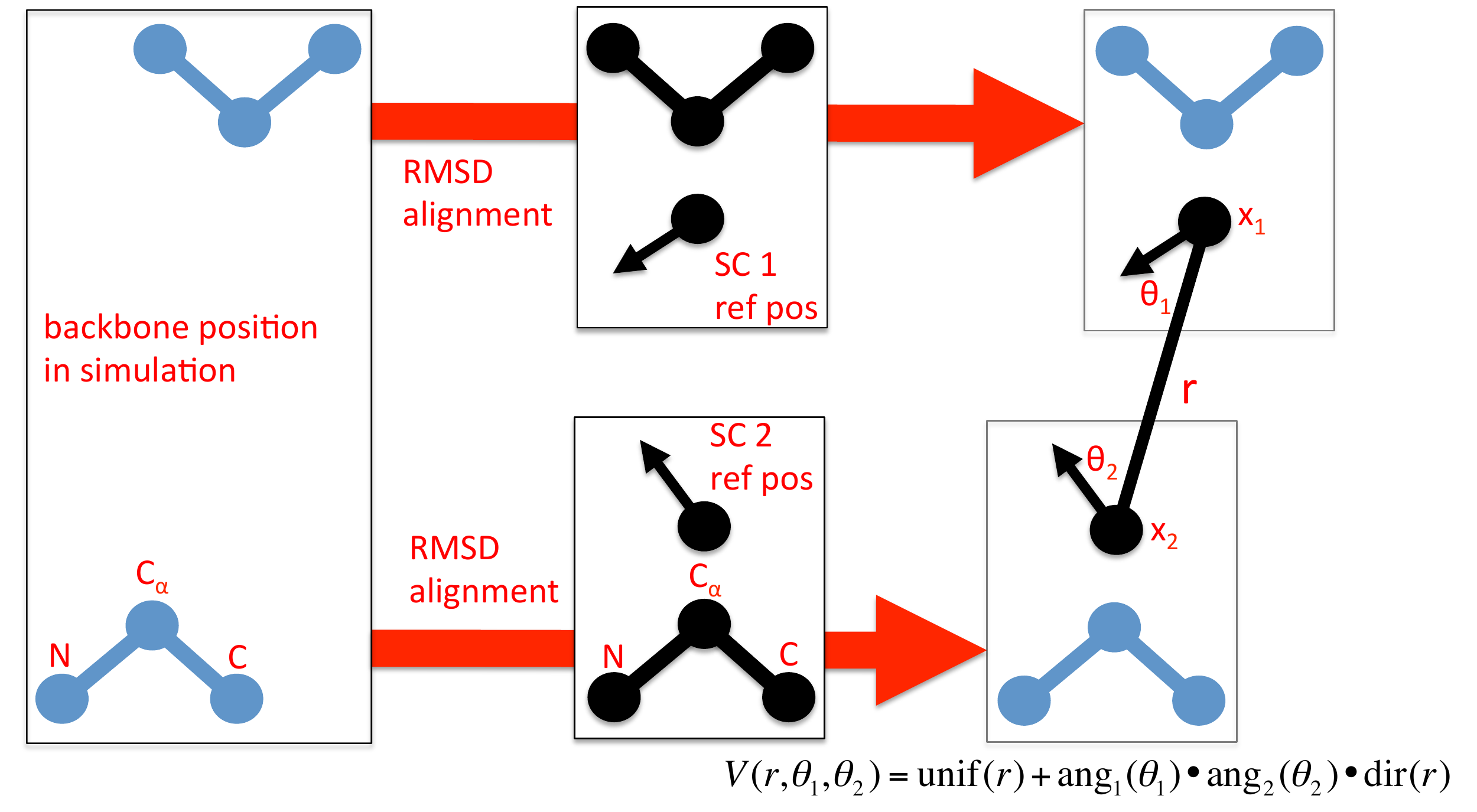}
    \caption{
	Coordinates used for side chain interactions.  For each residue, a
	reference backbone structure is aligned to the N, C$_\alpha$, and C
	atomic coordinates.  This alignment creates a reference frame to
	establish the position and direction of the side chain bead. The two
	side chain beads $x_1$ and $x_2$ for a pair of residues establishes
	three coordinates, the distance $r$ and angles $\theta_1$ and
	$\theta_2$.
    } 
\end{figure}

For side chain-side chain interactions the $\kappa$ prefactor is 1, but for side
chain-backbone interactions $\kappa$ depends on the hydrogen bonding state of
the backbone residue.  This distinction reflects that the presence of one
hydrogen bond inhibits the formation of another due to competition for the lone
pair of electrons. Specifically, the interaction between a backbone hydrogen or
oxygen is given a hydrogen bond confidence score $f$, a number that is typically
close to 0 for non-hydrogen bonded and 1 for hydrogen bonded residues. We set
$\kappa = 1-f$ so that the interaction is only turned on for hydrogens or
oxygens that are not participating in a backbone-backbone hydrogen bond.  The
physical motivation is that the directional interaction primarily describes the
effects of the dipole interactions, and in a hydrogen bond the C=O and N--H
dipoles have approximately zero total dipole moment.  While it is theoretically
possible for the algorithm to carefully balance hydrogen and oxygen interactions
that themselves cancel out on hydrogen bonded pairs, it is much easier to
achieve a physically reasonable model if we enforce the zeroing of directional
interactions with already hydrogen-bonded pairs.The hydrogen bond distance and
angular criteria are detailed in section \ref{sec:sim-details}. 

The side chain-backbone interactions are needed to describe helix capping, where
a side chain atom forms a hydrogen bond with an otherwise unsatisfied donor or
acceptor at the end of helix. We have observed that a proper description of
these capping effects is required to avoid helix fraying and inordinately long
helices. Harper and Rose\cite{harper1993helix} have observed that N-terminal
capping of a helix  by side chains is more likely to be observed than is
C-terminal capping by the side chain.  This finding is consistent with our
maximum-likelihood training (below), where side chain-amide hydrogen
interactions are fit with stronger (i.e.~higher confidence) potentials than side
chain-oxygen interactions.  Harper and Rose also note that hydrophobic residues
play a strong role in helix capping by covering the exposed protein backbone at
the ends of helices.  To provide our model with the freedom to describe this
effect, an additional side chain-backbone interaction is added with three beads
representing the hydrophobic portion of the backbone.  The location of the three
beads are initialized from the reference position of N, C$_\alpha$, and C and
are optimized with the rest of the parameters.  For this interaction, $\kappa =
1$. 

\section{\label{ref:max-lik} Maximum-likelihood training}

\subsection{Training objective function}

The side chain model is trained by the maximum-likelihood principle.
Specifically, we determine the set of parameters that maximizes the log
probability of the true side chain states $\cb_p$ in the Boltzmann ensemble of
all possible side chain states $\cb$ for the fixed backbone positions $X_p$ for
each protein $p$.

\begin{align}
     p(\cb_p) &= \frac{e^{-V(\cb_p)}}{\sum_{\cb} e^{-V(\cb)}} \\
    -\log p(\cb_p)&= V(\cb_p) + \log \left(\sum_{\cb} e^{-V(\cb)}\right) \\
	          &= V(\cb_p) - G^\text{SC} \label{eqn:truegap}\\
                  &= E_\text{gap}.
\end{align}
The evaluation of $E_\text{gap}$ requires the evaluation of the free energy of
the side chains, a quantity that is intractable to calculate exactly.
Fortunately, our side chain energy Eq.~\ref{eqn:free-energy} approximates the true
side chain free energy $G^\text{SC}$ that appears in Eq.~\ref{eqn:truegap}.
Furthermore, the expression for the parametric derivative
Eq.~\ref{eqn:ref-param-deriv} allows for gradient descent optimization to minimize
the average gap energy.  

\subsection{Training results}

The accuracy of the results are computed in two ways.  The first measure
computes the accuracy of the one-residue probabilities at predicting the
$\chi_1$ states of the protein.  This quantity is the traditional accuracy
measure for side chain packing algorithms.  The second measure is the quality of
the ensemble, obtained by computing the difference between the free energy of
the side chain system and the potential energy of the crystallographic rotamer
configuration.  For a highly accurate side chain ensemble, we would expect that
the crystal configuration would be a high probability state in the ensemble and
thus this energy gap would be small.  This energy gap is minimized by the
maximum-likelihood training.  The two accuracy measures are typically linearly
related for the side chain models we consider.

To compare to state-of-the-art side chain prediction methods, we compare to
SCWRL4\cite{scwrl4} on its training and validation set of side chains
conformations, as well as the RASP algorithm\cite{rasp} for rapid side chain
packing.  Since the Upside model lacks full side chains, we use the most likely
$\chi_1$ rotamer state according to the 1-residue marginal distributions
$p_i(\cb_i)$.  As per SCWRL4's validation procedure, the side chains with less
than 25$^\text{th}$ percentile electron density are excluded.  To avoid biasing
the comparison toward Upside, the SCWRL4 set of proteins is split so that 20\%
of the proteins are withheld for measuring accuracy, while the rest are used for
maximum-likelihood training of Upside. The accuracy metric chosen is to
calculate the fraction of side chains for which the Upside or SCWRL4 predicted
$\chi_1$ rotamer state agrees with the crystallographic conformation.  The
residues alanine, glycine, and proline are excluded from the comparison.

\begin{figure}
    \centering
    \includegraphics[width=0.7\linewidth]{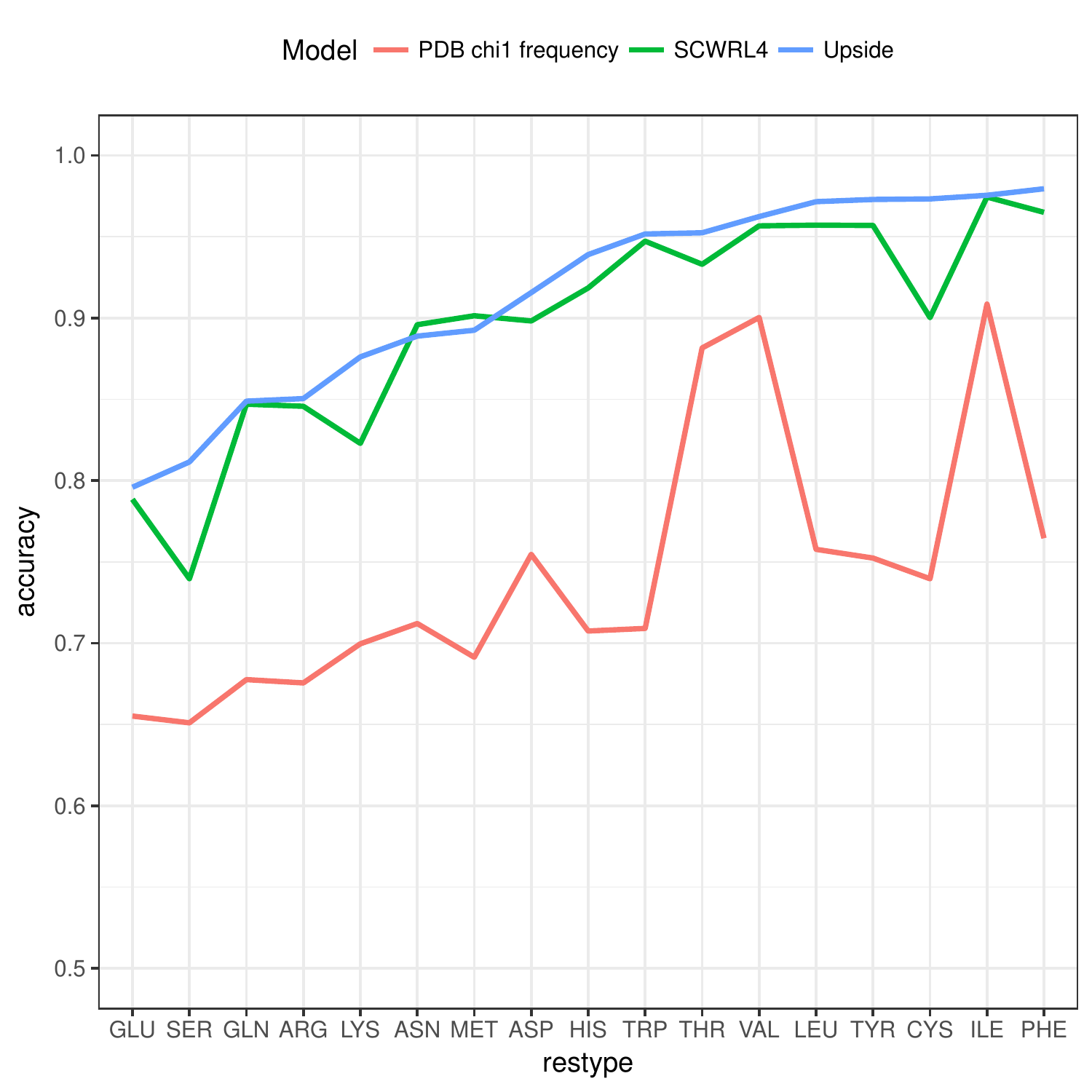}
    \caption{\label{fig:scwrl} Comparison of $\chi_1$ prediction accuracy for
    Upside and SCWRL4, ordered by Upside accuracy.  The ``PDB $\chi_1$
    frequency'' line represents the accuracy of the NDRD rotamer library without
    any interactions; this library is used in both Upside and SCWRL4.}
\end{figure}

\begin{table}
    \centering
    \begin{tabular}{lr@{}lr@{}l}
	Parameters & \multicolumn{2}{c}{Accuracy change (\%)} &
	\multicolumn{2}{c}{$\Delta E_\text{gap}$ ($k_\text{B}T$)} \\
	10\AA\ cutoffs                 &  +0&.7 & -0&.028 \\
	Full model                     &   0&.0 &  0&.000 \\
	No H/O interactions            &  -0&.6 &  0&.013 \\
	No N,C$_\alpha$,C beads        &  -2&.3 &  0&.040 \\
	$\phi,\psi$-independent $V(\chi)$     &  -3&.3 &  0&.004 \\
	Isotropic only                 &  -3&.5 &  0&.080 \\
	Repulsive only                 &  -3&.8 &  0&.060 \\
	Side chain--side chain only    &  -3&.8 &  0&.067 \\
	Side chain--backbone  only     &  -6&.1 &  0&.125 \\
	No interactions                & -13&.7 &  0&.435
    \end{tabular}
    \caption{\label{tbl:ablation} The significances of various components of the
    model reflect the decrease in accuracy for their removal.  The parameters
    are separately optimized for each row of the table so that each
    $E_\text{gap}$ represents the best achievable for the indicated functional
    form.  Note that these predictions are based on single-chain structures, so
    they differ slightly in accuracy from the predictions on all-chain
    structures reported in Figure \ref{fig:overall-accuracy}.}
\end{table}

\begin{figure}
    \centering
    \includegraphics[width=\linewidth]{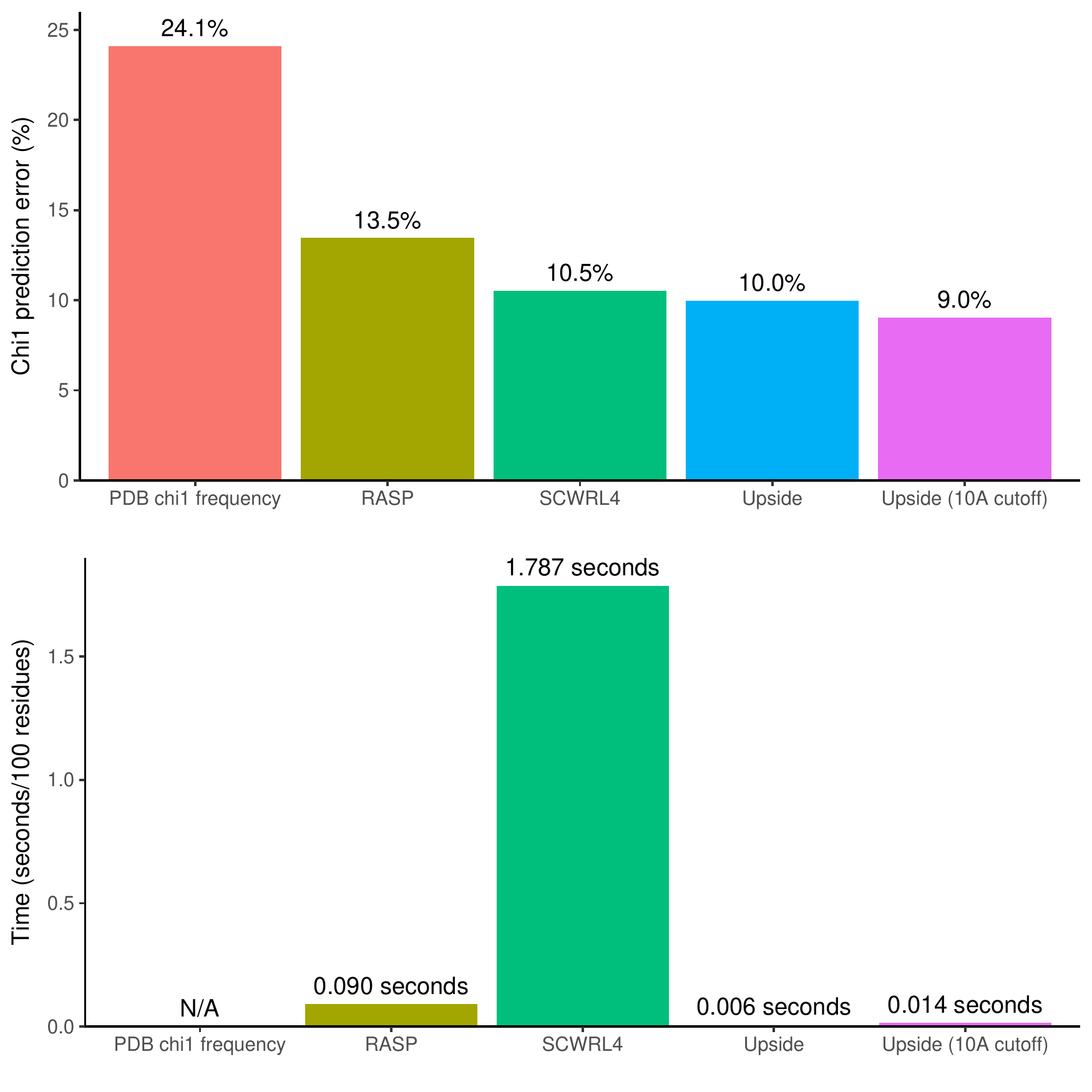}
    \caption{\label{fig:overall-accuracy} Comparison of accuracy of predicting
    side chains with high electron density as well as execution. For all
    programs, time spent reading the protein structure and writing the results
    is excluded from the execution time to focus on the cost of solving for the
    side chain positions.  For Upside (10 \AA\ cutoff), all side chain
    interactions with backbone or other side chains are cutoff at 10
\AA.}
\end{figure}

The importance of various interactions may be examined directly in Upside by
noting the change in accuracy upon their modification.  For example in Table
\ref{tbl:ablation}, we can see that using only repulsive interactions causes a
3.8\% drop in side chain prediction accuracy compared to the full model, which
quantifies the importance of the attractive interactions.

Upside is very accurate, predicting the correct $\chi_1$ rotamer 91.0\% of the
time using 10 \AA\ cutoffs, which is significantly better than either
SCWRL4\cite{scwrl4} or RASP\cite{rasp} (Fig \ref{fig:overall-accuracy}).
Additionally, Upside predicts side chains 16 times faster than the
speed-optimized RASP and 300 times faster than accuracy-optimized SCWRL4.  This
very fast performance enables Upside's side chain model to be viable in the
inner loop of molecular dynamics, as discussed in the next section.  

\section{Molecular dynamics}

The parameters obtained from the maximum-likelihood training are optimized for
side chain packing for a set of fixed, native backbones. In the limit that the
model is flexible enough to model the true side chain interactions and there is
unlimited training data, the maximum-likelihood method would recover the true
side chain interaction. Even without having the true form of the side chain
interaction, the maximum-likelihood parameters assign high probability to the
observed rotamer states, thereby including at least some of the underlying
physics.  

To test the suitability of adapting the side chain packing model to molecular
dynamics, simulations were run on small, fast-folding proteins.  To create a
reasonable protein dynamics model, backbone springs, backbone sterics, hydrogen
bond energy, and a basic Ramachandran potential were added to the side chain
model (section \ref{sec:sim-details}).  The Ramachandran potential is derived
from a coil library\cite{ting2010neighbor} as a statistical potential. The
hydrogen bond enthalpy is varied to find the maximum accuracy.  Note that
because alanine and glycine have no side chain rotamer states, and hence no
training to match the native $\chi$-angles can be conducted, the ALA-ALA,
ALA-GLY, and GLY-GLY potentials are completely determined by the regularization.
Interactions of ALA and GLY with other residue types are optimized, however, as
rotamer states of the other residues provide information on the ALA-X and GLY-X
interactions.

\begin{figure}
    \centering
    \includegraphics[width=\linewidth]{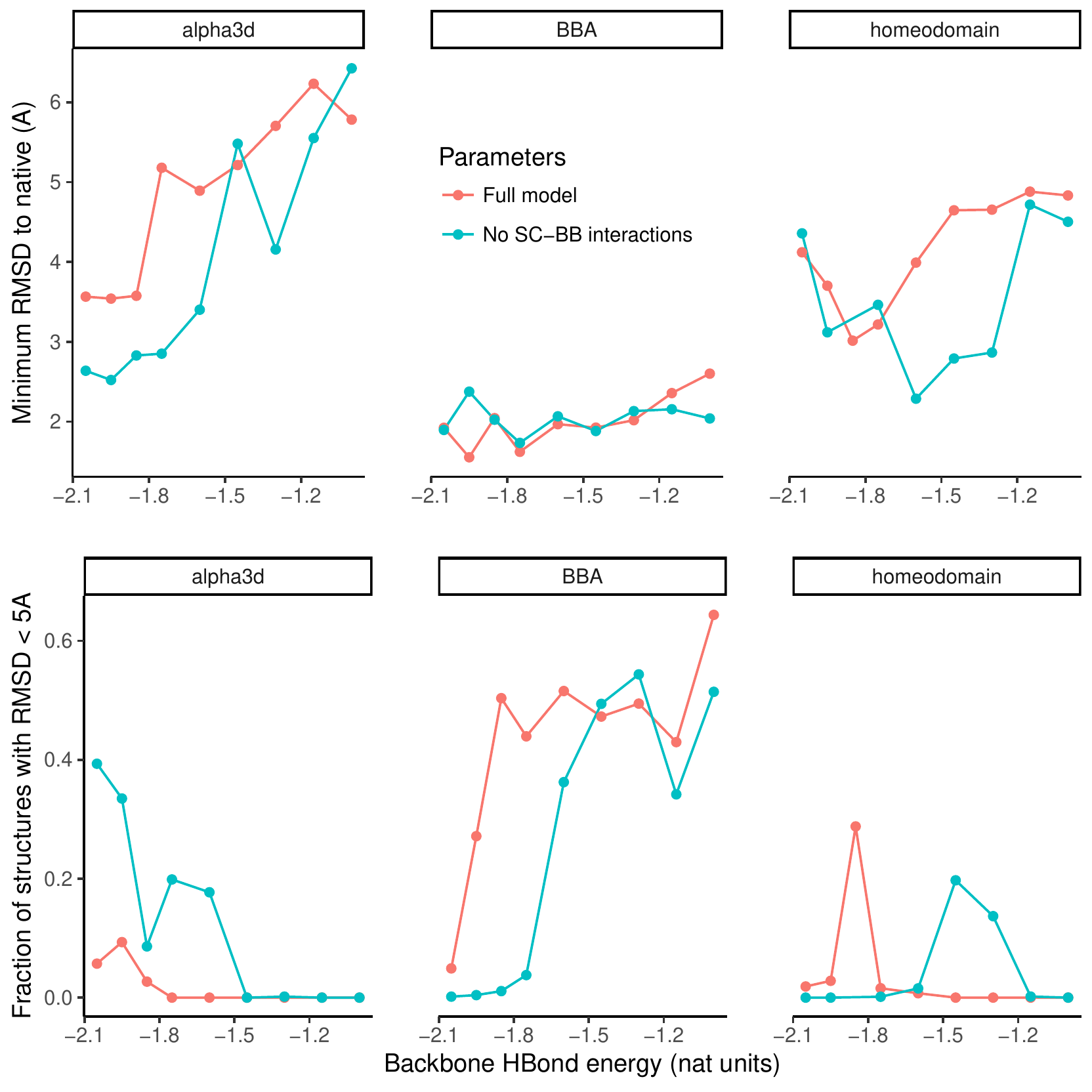} 
    \caption{ Accuracy of MD simulations for three proteins at variable backbone
    hydrogen bond strength.  This term does not play an explicit role in the
    packing optimization as the backbone is fixed. To assign this term an
    energy, it was manually varied for the best simulation accuracy.  This term
    is the only parameter  directly optimized for simulation accuracy.}
\end{figure}

\begin{figure}
    \centering
    \includegraphics[width=\linewidth]{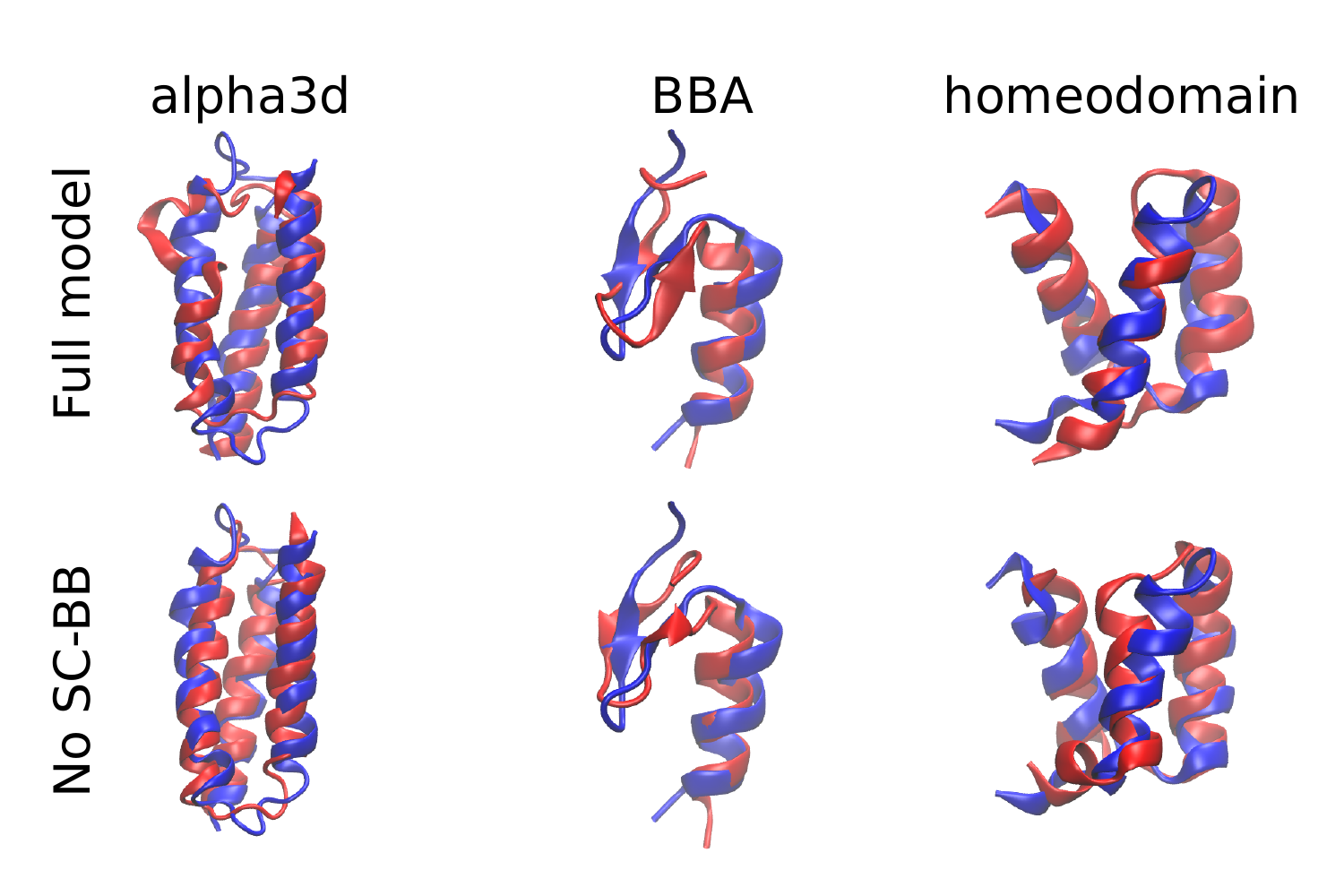} 
    \caption{ Closest structure to native for each protein and side chain model
    at optimal hydrogen bonding strength.  Blue is the native structure and red
    is the simulation.}
\end{figure}

\section{Related Work}

We highlight several works related to the major features of our model, including
molecular dynamics on three atoms but with a dynamic ensemble of side chains,
optimized discretization of the side chain states to best represent the protein
interactions in the coarse-grained model, a potential with optimized and
state-dependent bead locations and orientations, training a protein interaction
model for folding using side chain packing accuracy, and a side chain model with
an explicit side chain entropy.

A large body of work, exemplified by SCWRL4\cite{scwrl4}, has studied the
prediction of side chain configurations by discrete rotamer states.  SCWRL4
achieves approximately 90\% $\chi_1$ accuracy for predicting the most likely
rotamer states by minimizing the energy that combines observed rotamer state
frequencies and an atomic interaction model\cite{scwrl4}.  A variety of
algorithms have been developed for solving for the highest probability side
chain states given the pair interaction values\cite{desmet1992dead,xu2006fast}.
Kamisetty et al.\cite{kamisetty2008free} have worked on scoring protein
interaction complexes using a self-consistent approximation to the side chain
interactions.  Earlier simulation work by Koehl and
Delarue\cite{koehl1994application} use 1-residue mean field techniques to
approximate ensembles of side chain conformations but fail to account for the
pairwise correlations of the side chain rotamer states.  All of these works use
atomically-detailed descriptions of the side chains paired with simple or
molecular dynamics interaction terms.  Their highly detailed side chains with
many $\chi$-angles for each residue make it difficult to perform calculations
sufficiently fast for folding simulations, and the use of existing interactions
(instead of a newly-trained interaction model) makes it difficult to reduce
detail to speed computation.  There has also been extensive work in
reconstructing backbone positions from side chain beads\cite{backbone-from-sc}
in lattice models, but these models do not perform a proper summation over
possible rotamer states.

RASP\cite{rasp} is side chain modeling program designed to significantly improve
the speed of side chain packing while achieving comparable accuracy.  The
authors use careful selection of the most important energy terms as well as
employing clash-detection to guide the optimization of the side chain
conformations.

There have also been a large number of coarse-grained techniques that use a
variety of non-isotropic potentials for reduced side chain interactions.  One of
the most successful is the coarse-grained united residue model
(UNRES)\cite{liwo2000unres}.  The model also uses statistical frequencies to
determine the positions of the side chains but it emphasizes the
parameterization of the coarse-grained model from physics-based calculations
instead of statistical information. Though the  potential form (Gay-Berne) used
in UNRES is quite different from our work, UNRES also uses non-isotropic side
chain potentials\cite{unres-backbone}.
 
Similar to our work, Dama, Sinitskiy, et al.\cite{dama2013theory} investigate
mixed continuous-discrete dynamics, where the states of molecules jump according
to a discrete Hamiltonian.  Their method differs from our work in a number of
important ways: the authors use discrete jumps in state instead of a free energy
summation over all states that we employ; they do not optimize the rotamer
states as we do; and they train parameters from force matching of molecular
dynamics trajectories rather than from the statistical analysis of experimental
data as we employ.

\section{Discussion and conclusion}

We have demonstrated a fast, principled method to coarse-grain discrete side
chain states to create a smooth backbone potential.  This procedure results in a
considerable decrease in computational time as it removes the side chain
rattling and friction normally associated with a polypeptide chain moving in a
collapsed state. This tracking and instantaneous equilibration of the side
chains is analogous to the instantaneously-equilibrated electronic degrees of
freedom with respect to the nuclear motions employed in the adiabatic
Born-Oppenheimer approximation\cite{born1927quantentheorie}. Motions are
calculated only for three heavy backbone atoms, yet the model contains
considerable structural detail including hydrogen bonds involving both the
backbone and side chains. Further, we have demonstrated both a maximum
likelihood procedure to obtain a physically-reasonable potential from the side
chain packing of X-ray structures and a tunable discretization of the rotamer
states.  The resulting method is capable of rapid molecular dynamics sampling of
protein structures with moderate accuracy.

The reason that Upside is both faster and more accurate than competing methods
at side chain packing is that it shifts the complexity of the
$\chi_1$-prediction problem.  Traditional side chain prediction uses a detailed
configuration space of all rotamers and side chain atoms but simple interaction
forms with few parameters.  Upside uses a coarse configuration space with only a
single directional bead per residue but a complex and well-optimized set of
parameters consisting of over 10,000 jointly-optimized parameters (trained on
approximately 500,000 residues).  Upside demonstrates that $\chi_1$ rotamers can
be predicted with state-of-the-art accuracy without needing to examine
fine-grained atomic packing.  Additionally, the side chains in Upside are
represented as a Boltzmann ensemble whose 1-residue marginal probabilities are
used to predict $\chi_1$ instead of predicting $\chi_1$ using the lowest energy
configuration.  This approach allows for the natural consideration of side chain
entropy and conformational variability.  Creating a Boltzmann ensemble over
rotamer states also allows exact, continuous forces to be defined for the
approximate ensemble, enabling molecular dynamics using potential energies
already validated to represent the physics of side chain packing.

A natural question is whether the strengths of SCWRL4 and this algorithm may be
combined.  There are two reasons to believe that such a combination would be
fruitful.  The first reason is that when Upside and SCWRL4 predict the same
$\chi_1$ rotamer, the prediction is 95.4\% accuracy, substantially more accurate
than either program alone.  This suggests Upside and SCWRL4 provide independent
information about the side chain conformations and hence, combining both
approaches should produce a substantially better packing model.    The second
reason that Upside and SCWRL4 may be combined is that Upside provides
probability functions as its outputs, rather than just the minimum energy
conformation as in SCWRL4.  The underlying SCWRL4 single-rotamer energies could
be augmented with $-\lambda \log p_\text{upside}(\cb)$.  For an appropriately
determined $\lambda$, this should incorporate some of Upside's information
directly into SCWRL4, increasing SCWRL4's accuracy.  Alternatively, SCWRL4's
detailed but simple energy function could be augmented by an Upside-style
coarse-grained function, possibly with additional maximum-likelihood tuning.

The importance of optimizing the bead locations and directions in our model
illustrates the principle that chemical intuition can only be a partial guide to
accurate coarse-graining of protein interactions.  The location of the
interaction sites has a strong effect on our model's ability to achieve
high-packing accuracy, and we expect similarly strong effects would be observed
had we directly optimized for backbone conformational accuracy.

For dynamics applications, Upside shows promise as a route to accurate and
inexpensive molecular simulation but requires further development.  New training
techniques are being developed to directly optimize the backbone accuracy of the
Upside model.  Preliminary results from the new training methods indicates that
we are able to achieve dramatic improvements in the accuracy of \textit{de novo}
folding while preserving the rapid folding properties.  We expect that our
belief-propagated side chains will serve as an excellent basis for new methods
in protein simulations.

\section{Author Contributions}

J.M.J.~invented algorithm, designed research, performed research, analyzed results, and wrote
manuscript. K.F.F.~designed research and wrote manuscript.  T.R.S.~designed
research, analyzed results, and wrote manuscript.

\begin{acknowledgments}
This research is supported, in part, by National Science Foundation (NSF) Grant
No.~CHE-1363012, NIH grant GM 55694, and by the NIGMS of the National
Institutes of Health under award number T32GM008720.  This work was
completed in part with resources provided by the University of Chicago
Research Computing Center.

We would like to thank Sheng Wang and Jinbo Xu for helpful discussions during
this research.  Zhichao Miao kindly provided a modified version
of RASP to measure only the side chain packing time.
\end{acknowledgments}

\bibliography{master}

\setcounter{section}{0}
\renewcommand{\thesection}{Appendix \Alph{section}}
\section{\label{apx:beliefs} Belief propagation}
For convenience, this appendix contains a brief description of the equations
used to implement belief propagation for the side chain free energies.  Given
1-residue energies $v_i(\cb_i)$ and 2-residue energies $v_{ij}(\cb_i,\cb_j)$, we
seek probabilities $p_i(\cb_i)$ and $p_{ij}(\cb_i,\cb_j)$ to minimize the free
energy in Eq.~\ref{eqn:free-energy}.

It is helpful to first understand the intuition behind the belief propagation
process.  We seek a consistent set of 1- and 2-side chain probabilities for the
residues compatible with the interaction potential Eq.~\ref{eqn:ising}.  The
probability of each residue state $\cb_i$ for residue $i$ is determined by two
factors.  The first factor is the 1-residue energy $v_i(\cb_i)$ that would
determine the probabilities exactly in the absence of interactions.  The second
factor is consistency with the side chain states of the residues in contact with
residue $i$, where consistency is determined by the potentials
$v_{ij}(\cb_i,\cb_j)$.  
Belief propagation optimizes these factors to minimize the approximate free energy
Eq.~\ref{eqn:free-energy} as derived in reference
\citenum{yedidia2003understanding}.  The iteration is described more formally
below, including a damping term $\lambda$ to suppress oscillations during the
self-consistent iteration.

For 1-residue beliefs, define $b_i^r(\cb_i)$ to be the round $r$ ``belief'' that
the $i$-th residue is in state $\cb_i$.  For the 2-residue beliefs, we have two
beliefs for each pair of interacting residues (i.e.~any pair of residues that
interact in any rotamer states).  Define $b_{ij}^r(\cb_j)$ to
be the round $r$ belief for the residue pair ($i$,$j$) that residue $j$ is in
state $\cb_j$.  The belief $b_{ji}^r(\cb_i)$ is defined similarly.  

To initialize the algorithm at round 0, we take 
\begin{align}
    b_i^0   (\cb_i) &= e^{-v_i(\cb_i)} \\
    b_{ji}^0(\cb_i) &= \sum_{\cb_j} e^{-v_{ij}(\cb_i,\cb_j)} b_j^0(\cb_j).
\end{align}
We compute the round $r+1$ beliefs from the round $r$ beliefs according to the
following equations.
\begin{align}
b_{ji}^{r+1}(\cb_i) &= 
    \sum_{\cb_j} e^{-v_{ij}(\cb_i,\cb_j)} \frac{b_j^r(\cb_j)}{b_{ij}^{r}(\cb_j)}
    \label{eqn:update_one} \\
b_i^{r+1}(\cb_i) &= \lambda b_i^r(\cb_i) + (1-\lambda)
    \frac{         e^{-v_i(\cb_i)} \prod_j b_{ji}^{r+1}(\cb_i)}
    {\sum_{\cb_i} e^{-v_i(\cb_i)} \prod_j b_{ji}^{r+1}(\cb_i)} \label{eqn:update_two}
\end{align}
The products in Eq.~\ref{eqn:update_one} should be understood as taken
only over residues $j$ that interact with residue $i$.  The damping constant
$\lambda$ suppresses oscillatory behavior that hinders convergence
($\lambda=0.4$ is used in the present work). The equations are iterated until
$|b_i^{r+1}(\cb_i)-b_i^r(\cb_i)|<0.001$ for all residues $i$ and states
$\cb_i$.

From the converged beliefs $b_i(\cb_i)$ and $b_{ij}(\cb_j)$, we can compute
the marginal probabilities
\begin{align}
    p_i(\cb_i) &= b_i(\cb_i) \\
    p_{ij}(\cb_i, \cb_j) &=
    \frac{
        \frac{b_i(\cb_i)}{b_{ji}(\cb_i)}
        e^{-v_{ij}(\cb_i,\cb_j)}
        \frac{b_j(\cb_j)}{b_{ij}(\cb_j)}
    }{
        \sum_{\cb_i,\cb_j}
        \frac{b_i(\cb_i)}{b_{ji}(\cb_i)}
        e^{-v_{ij}(\cb_i,\cb_j)}
        \frac{b_j(\cb_j)}{b_{ij}(\cb_j)}
    }.
\end{align}
The free energy of the model is obtained by using the marginal probabilities
above in Eq.~\ref{eqn:free-energy}.


\newpage

(SUPPLEMENTAL INFORMATION ON NEXT PAGE)

\clearpage

\begin{widetext}
\begin{center}
    \textbf{\large Supplemental Information}
\end{center}
\end{widetext}
\beginsupplement

\section{Training set}

The side chain packing interaction is trained using a large, non-redundant
collection of crystal structures from the PDB with 50--500 residues and
resolution less than 2.2\AA.  From a training set of protein structures, we
extract the sequences $s_p$, backbone trace positions $X_p$, and  true
coarse-grained side chain states $\cb_p$ for each protein $p$.  The proteins are
further filtered using PISCES\cite{wang2003pisces} so that all pairs of proteins
have sequence similarity less than 30\%.  Non-globular structures in the dataset
are removed, as we suspect that the side chain packing of these structures is
more strongly influenced by other chains in the crystal structures.  We define
non-globular structures as outliers in the linear relationship between
$\log(N_\text{res})$ and $\log(R_g)$; the outliers are identified using the
RANSAC algorithm\cite{fischler1981random}.  After filtering, 6255 chains
remained, containing approximately 1.4 million residues. 

\section{\label{sec:sim-details} Simulation details}

\begin{table}[h]
    \centering
\tiny
\begin{tabular}{llll}
Name & PDB ID & Length & Sequence \\
alpha3d & 2a3d & 73 & MGSWAEFKQRLAAIKTRLQALGGSEAELAA \\
        &      &    & FEKEIAAFESELQAYKGKGNPEVEALRKEA \\
	&      &    & AAIRDELQAYRHN \\
BBL & 2wxc & 47     & GSQNNDALSPAIRRLLAEWNLDASAIKGTG \\
	&      &    & VGGRLTREDVEKHLAKA \\
homeodomain & 2p6j & 52 & MKQWSENVEEKLKEFVKRHQRITQEELHQY \\
	&      &    & AQRLGLNEEAIRQFFEEFEQRK \\
\end{tabular}
    \caption{Sequences of proteins for molecular dynamics}
\end{table}

All simulations are run with Upside, a custom simulation engine that implements
the belief propagation of side chain interactions as well as the parameter
derivatives needed for gradient descent.

The replica exchange temperatures are 0.500, 0.532, 0.566, 0.600, 0.636, 0.672,
0.709, 0.748, 0.787, 0.828, 0.869, 0.912, 0.955, and 1.000.  The Ramachandran
potential uses the NDRD TCB coil library\cite{ting2010neighbor}.  The backbone
hydrogen bond interaction uses both distance and angle criteria to determine
hydrogen bonds.  The H-O bond distance interaction starts at approximately
1.4\AA\ and ends at 2.5\AA.  Both the N-H-O and H-O-C criteria half-heights are
at approximately 47 degrees off of collinear.

We use Verlet integration with a time step of 0.009 units.  We use the random
number generator Random123 \cite{salmon2011parallel} to implement the Langevin
dynamics with a thermalization time scale of 0.135 time units.  The
thermalization time scale (related to Langevin friction) is chosen to maximize
the effective diffusion rate of chains while effectively controlling simulation
temperature.  As Langevin dynamics with any friction coefficient produces the
same Boltzmann ensemble, we chose to maximize equilibration of our system rather
than attempt to match a solvent viscosity.

The cutoff radius for side chain-side chain interactions is 7\AA, and the cutoff
radius for side chain-backbone interactions is 5\AA.  The distance splines are
zero-derivative-clamped cubic splines with a knot spacing of 0.5\AA.  The
angular splines have a knot spacing of 0.167 in $\cos \theta$, which ranges over
$[-1,1]$.

\section{Optimization details}

The Adam optimizer\cite{kingma2014adam} is used to minimize the energy gap.
This optimizer is convenient because it automatically adjusts the gradient
descent step size for each parameter according to the typical scale of the
gradient in that dimension.  This rescaling is important because spline
coefficients at large radii tend to have much larger gradient magnitudes than
parameters at small radii. 

We use the following settings for the Adam optimizer: minibatch size 256
proteins, $\alpha=0.03$, $\beta_1=0.90$, $\beta_2=0.96$, $\epsilon=10^{-6}$.
Positivity constraints on the angular coefficients are enforced by a exponential
transform.  The regularization integrals over all space are approximated by sums
at the knot locations of the radial and angular splines.

A regularization penalty is added to the maximum-likelihood optimization that
encourages smoothness of the potential.  This penalty also reduces the
validation error of the training.  The regularization penalties chosen are
\begin{align}
    &\sum_i (2 c_i^\text{unif} - c_{i-1}^\text{unif} - c_{i+1}^\text{unif})^2
    \label{eqn:lap} \\
    &\sum_i (c_i^\text{dir})^2 \label{eqn:dir-min} \\
    &\sum_i (c_0^\text{unif} - (5\,k_\text{B}T))^2 \label{eqn:origin-penalty}
\end{align}
The penalty Eq.\ref{eqn:lap} encourages a small second derivative for the
isotropic ($\unif$) term, while the penalty Eq.\ref{eqn:dir-min} minimizes the
size of the directional interactions.  Finally, the penalty
Eq.\ref{eqn:origin-penalty} ensures a strong steric core for interactions.

The derivative calculations needed for regularization and coordinate transforms
are handled with the Tensorflow framework\cite{tensorflow2015whitepaper}.

\section{Optimized mapping to coarse states}\label{sec:state-decomposition} 

The $\chi$-angles for the side chains are partitioned into discrete states in an
optimized manner. The NDRD rotamer library\cite{shapovalov2011smoothed} provides
a set of approximate discrete states for each residue type according to their
frequencies of occurrence in a non-redundant set of high resolution protein
structures in the PDB. However, the number of rotamer states in the NDRD library
can be quite large.  For instance, naively using all 81 rotamers for each
arginine means that computing the pair interaction $v_{i,j}$ for two arginines
would require computing $81^2 = 6561$ energy values.  Consequently, instead of
using all possible rotamer states, several NDRD rotamer states are combined into
3--6 coarse-grained rotamer states for the sake of manageable computational
cost.

We choose to aggregate the rotamer states of the side chain to minimize the
positional uncertainty of side chain atoms in each state.  A search over all
possible aggregations is conducted to find the aggregation that provides the
smallest possible error.  More formally, the NDRD rotamer
library\cite{shapovalov2011smoothed} is used to define the atomic positions
$x_{ij}^f(\phi,\psi)$, where $i$ is the atom (such as C$_\beta$), $j$ is the
coordinate ($x$, $y$, or $z$), and $f$ is the fine-grained rotamer state.  Each
rotamer state has a probability $p^f(\phi,\psi)$ specified in the NDRD library
from frequencies in the PDB for each fine-grained rotamer state as a function of
the backbone dihedral angles $(\phi,\psi)$.  Each fine-grained state $f$ may
belong to exactly one coarse-grained state $c$ (i.e.~the $c$ states form a
partition of the $f$ states). Given the choice of a coarse-grained state $c$, an
average is performed over the fine-grained atomic positions, and sum is taken
over the probabilities of all fine-grained states $f$ grouped into $c$ according
to the prescription,
\begin{align}
    q^c(\phi,\psi) &= \sum_{f \in c} p^f(\phi,\psi) \\
    y_{ij}^c(\phi,\psi) &= \frac{1}{q^c(\phi,\psi)} \sum_{f \in c} p^f_{ij}(\phi,\psi)\,
    x^f_{ij}(\phi,\psi),
\end{align}
where $q^c$ is the coarse-grained probability and $y_{ij}^c$ is the
coarse-grained atomic position.

The error incurred by coarse-graining is defined as the variance of the atom
positions within each coarse-grained state, weighted by the frequency of
occurrence of the coarse-grained state in the PDB.  Specifically, the error
$\sigma^2(\phi,\psi)$ is defined as,
\begin{align}
    \sigma^2(\phi,\psi) = \sum_f \frac{p^f(\phi,\psi)}{N_\text{atom}} \sum_{ij} 
    (x_{ij}^f(\phi,\psi) - y_{ij}^{c(f)}(\phi,\psi))^2,
\end{align}
where $N_\text{atom}$ is the number of atoms in the side chain and $c(f)$ is the
coarse-grained state $c$ that contains the fine-grained state $f$.  The error
depends implicitly on the state decomposition $c(f)$ and measures the deviation
of the atoms within each state. This error favors the fine-grained states $f$
that occur with higher frequency in the PDB.

The division of fine-grained states into coarse-grained states is restricted for
simplicity  to be independent of the Ramachandran angles for the residue,
\begin{align}
    \sigma^2 = \int p^\text{Rama}(\phi,\psi)\, \sigma^2(\phi,\psi)\, d\phi\, d\psi,
\end{align}
where $p^\text{Rama}(\phi,\psi)$ is the frequency of each Ramachandran angle taken
from the PDB coil library.  Note that this error term depends implicitly on the
decomposition $c(f)$ and weights for the  $(\phi,\psi)$ pairs according to their
frequencies in the coil library.

An optimal coarse-grained representation of the side chain rotamer states is
obtained by minimizing $\sigma^2$ for each residue type over all partitions
$c(f)$.  We force the coarse-graining $c(f)$ to obey a few conditions,
essentially to make sure that $c(f)$ is easily interpretable in terms of
$\chi_1$ and $\chi_2$ as well as limiting the number of possibilities that must
be checked by the brute-force minimization.  In particular, the mapping from
coarse states back to $\chi_1$ rotamer states is unambiguous because no single
coarse state contains two different $\chi_1$ rotamer states.  We impose the
following conditions:
\begin{enumerate}
    \item $c(f)$ depends only on the $\chi_1$ and $\chi_2$ rotamer states of $f$
	(i.e.~if $f_1$ and $f_2$ states differ only in their $\chi_3$ or
	$\chi_4$ states, then $c(f_1) = c(f_2)$).
    \item Each coarse state $c$ must contain only a single $\chi_1$ state 
    but may contain multiple distinct $\chi_2$ states for that $\chi_1$ state.
    \item Each coarse state $c$ must contain a contiguous range of $\chi_2$
	values.  This greatly reduces the number of possible coarse-grainings
	for residues with non-rotameric $\chi_2$ angles like asparagine.
\end{enumerate}
We optimize the decomposition of the coarse-grained state $c(f)$ by completely
enumerating all possible decompositions into coarse-grained states that satisfy
the three conditions above and contain no more than six coarse states.

\end{document}